\documentclass[11pt]{article}
\usepackage[utf8]{inputenc}
\usepackage{hyperref}
\usepackage{graphicx}
\usepackage{dcolumn}
\usepackage{bm}
\usepackage{longtable}
\usepackage[export]{adjustbox}
\usepackage{psfrag}

\usepackage{amsmath}
\RequirePackage{wasysym} \RequirePackage{setspace}\textheight=650pt
\pdfoutput=1
\begin{document}
\title{ {\bf Excited collective states of nuclei within Bohr Hamiltonian with Tietz-Hua potential
 }}
\author{M. Chabab$^{*,a}$, A. El Batoul $^{*,b}$, M. Hamzavi $^{\dag,c}$, A. Lahbas$^{*,d}$, M. Oulne$^{*,e}$ \\
\\ {\small $^*$ High Energy Physics and Astrophysics Laboratory, Faculty of Science Semlalia, }\\
{\small Cadi Ayyad University, P.O.B. 2390, Marrakesh,Morocco} 
\\ {\small $^{\dag}$ Department of Physics, University of Zanjan, P.O. Box 45195-313,
Zanjan, Iran }\\ \\
{\small $^{a}$ mchabab@uca.ma} \\
{\small $^{b}$ elbatoul.abdelwahed@edu.uca.ma}\\
{\small $^{c}$ majid.hamzavi@gmail.com} \\
{\small $^{d}$ alaaeddine.lahbas@edu.uca.ma} \\
{\small $^{e}$ Corresponding author : oulne@uca.ma} \\
}

\maketitle

\begin{abstract}
In this paper, we present new analytical solutions of the Bohr Hamiltonian problem that we derived with the Tietz-Hua potential, here used for describing the $\beta$-part of the nuclear collective potential plus harmonic oscillator one for the $\gamma$-part. Also, we proceed to a systematic comparison of the numerical results obtained with this kind of $\beta$-potential with others which are widely used in such a framework as well as with the experiment. The calculations are carried out for energy spectra and electromagnetic transition probabilities for $\gamma$-unstable and axially symmetric deformed nuclei. In the same frame, we show the effect of the shape flatness of the $\beta$-potential beyond its minimum on transition rates calculations. 
\end{abstract}
\section{Introduction}
The theoretical study of excited collective states in nuclei is of particular interest in nuclear structure inasmuch as it allows to understand shape phase transitions in nuclei. Therefore, different  approaches have been developed in this context particularly in the framework of Bohr-Mottelson model \cite{bohr,boh75} and Interacting Boson Model (IBM) \cite{iachello1987IBM}. Moreover, the interest devoted to such a thematic has increased even more with the occurrence of critical point symmetries. The pioneer symmetries were the E(5) \cite{E5} which has been introduced to describe phase transition between vibrational and $\gamma$-unstable nuclei and the X(5) \cite{X5} one elaborated with the aim of describing the phase transition between vibrational and axially symmetric prolate deformed nuclei.  Both critical point symmetries have used the infinite square well (ISW) as a recall potential for $\beta$-vibrations, while the $\gamma$-oscillations have been treated by the harmonic oscillator around $\gamma \approx 0$. Thereafter, the X(5) symmetry has generated the X(3) one \cite{bonatsos2006x} for $\gamma$-rigid nuclei and these two latter have been recently improved by the new X(3)-ML and X(5)-ML symmetries which have been elaborated by introducing for the first time the minimal length concept in nuclear structure \cite{chabab2016gamma}. Thus, the critical point symmetries have generally paved the way for the construction of other models by making use of different potentials leading to new exactly separable models allowing the description of nuclei which are near or far from the above mentioned critical point symmetries. One can cite for example the ES-E(5)  and the ES-X(5) models. The most popular among the used model potentials one can find Morse \cite{Inci11,bozt8,Inci15}, Kratzer \cite{Fortunato:2005dn,bonat13}, Davidson \cite{bonat11,bonat7,buganuenergy,CLO15P}, Sextic \cite{levai2004sextic,levai2010search,buganu15sextic,buganu12,buganu2011,Raduta:2013db,Buganu:2013}, Hulthén \cite{CLO15H}, Woods-Saxon \cite{capak15}, and Manning-Rosen \cite{CLO15MR,chabab2016electric} potentials. Recently, it has been shown that the Morse potential \cite{Inci11} is more appropriate than the Davidson one in the limit of E(5) and X(5) symmetries due to its shape which becomes flat when the variable $\beta$ increases in respect to the Davidson potential which grows as $\beta^2$. As it is known, the Morse potential has been introduced for the first time in molecular physics to describe the vibrational behavior of diatomic molecules. Furthermore, different forms of such a potential have been developed like for example Tietz-Hua potential \cite{tietz1963potential,hua1990four}. This latter is considered as more realistic than the Morse one in the description of molecular dynamics as moderate and high rotational and vibrational quantum numbers \cite{hamzavi2012rotation}. Thus, in the present work, we adopted this potential within the Bohr Hamiltonian for the $\beta$-part of the separable nuclear collective potential which is chosen in the following form : $v(\beta,\gamma)=u(\beta)+w(\gamma)/\beta^2$, while the $\gamma$-part $u(\gamma)$ of this latter is taken to be equal to a harmonic oscillator potential. The elaborated model in this work has been applied to study $\gamma$-unstable and axially symmetric prolate deformed nuclei through their energy spectra and transition probabilities. These nuclear characteristics have been obtained in closed analytical form by means of Nikiforov-Uvarov method \cite{NU}. The obtained numerical results in this work are in an excellent agreement with the experimental data and fairly better than those obtained with other potentials particularly for transition rates.

The paper is organized as follows. In the Section 2, the analytical expressions for the energy levels and excited-state wave functions are presented in the $\gamma$-unstable and prolate axial rotor case,  while the $B(E2)$ transition probabilities are given in Section 3.
 The numerical results for energy spectra and $B(E2)$ are presented, discussed, and compared with experimental data and available other models in Section 4. Finally, Section 5 contains our conclusion.
\section{Theory of the model}\label{S2}

 The original collective Bohr Hamiltonian is \cite{boh75}
\begin{equation}\label{eq1}
H =-\frac{\hbar^2}{2B}\left[\frac{1}{\beta^4}\frac{\partial}{\partial\beta}\beta^4\frac{\partial}{\partial\beta}+
\frac{1}{\beta^2\sin3\gamma}\frac{\partial}{\partial\gamma}\sin3\gamma\frac{\partial}{\partial\gamma}
-\frac{1}{4\beta^2}\sum_{k=1,2,3}\frac{Q^2_k}{\sin^2(\gamma-\frac{2}{3}\pi k)}\right]+V(\beta,\gamma),
\end{equation}

where $\beta$ and $\gamma$ are the usual collective coordinates, $Q_k$ are the components of angular momentum in the intrinsic frame, and $B$ is the mass parameter.
\subsection{$\gamma$-unstable case}
In order to achieve exact separation of the variables $\beta$ and $\gamma$ in eq. \eqref{eq1}, we choose the total wave function in the form \cite{wilet}
\begin{equation}\label{eq2}
\Psi(\beta,\gamma,\theta_i)=\xi(\beta)\Phi(\gamma,\theta_i)
\end{equation}
where $\theta_i(i=1,2,3)$ are the Euler angles, and we assume the potential to be $\gamma$ independent, $V(\beta,\gamma)=U_1(\beta)$.

Then, separation of variables leads to two equations: one depending only on the $\beta$ variable and the other depending on the $\gamma$ variable and the Euler angles,
\begin{equation}\label{eq3}
\left[-\frac{1}{\beta^4}\frac{\partial}{\partial\beta}\beta^4\frac{\partial}{\partial\beta}+\frac{\Lambda}{\beta^2}+u_1(\beta)\right]
\xi(\beta)=\epsilon\xi(\beta)
\end{equation}
and
\begin{equation}\label{eq4}
\left[-\frac{1}{\sin3\gamma}\frac{\partial}{\partial\gamma}\sin3\gamma\frac{\partial}{\partial\gamma}
+\frac{1}{4}\sum_{k}\frac{Q^2_k}{\sin^2(\gamma-\frac{2}{3}\pi k)}\right]\Phi(\gamma,\theta_i)=\Lambda\Phi(\gamma,\theta_i),
\end{equation}
where $\Lambda$ is the separation constant and the following notations are used $u_1(\beta)=\frac{2B}{\hbar^2}U_1(\beta)$ and $ \epsilon=\frac{2B}{\hbar^2}E$.

The $\gamma$ and Euler angles equation \eqref{eq4} has been solved by B\`es\cite{bes}. In this equation, the eigenvalues of the second-order Casimir operator SO(5) are expressed in the following form $\Lambda= \tau(\tau + 3)$, where $\tau$ is the seniority quantum number, characterizing the irreducible representations of SO(5) and taking the values $\tau=0, 1, 2, ...$ \cite{rakavy}. 

The values of angular momentum L occurring for each $\tau$ are provided by a well known algorithm and are listed in \cite{iachello1987IBM}. The ground state band levels are determined by $L=2\tau$.

\subsection{The prolate axial rotor case}
In this case, the reduced potential $v(\beta,\gamma)=2BV/\hbar^2$ depends on the asymmetry $\gamma$. However, to achieve exact separation of variables \cite{buganu2015analytical,bonat13,bonatsos2007,bozt8,wilet}, we assume a potential of the form $v(\beta,\gamma)=u_1(\beta)+u_2(\gamma)/ \beta^2$ , given in Eq. \eqref{eq1}.

For the $\gamma$-potential, we use a harmonic oscillator \cite{X5,bonatsos2007}
\begin{equation}
 u_2(\gamma ) =(3c)^2\gamma^2 \label{19}
\end{equation}
where $c$ is a free parameter.

As the $\gamma$ potential is minimal at $\gamma=0$, one can write the angular momentum term of Eq. \eqref{eq1} as \cite{X5}
 \begin{align}
\sum_ {k=1,2,3}\frac{Q_{k}^{2}}{\sin^2(\gamma-\frac{2}{3}\pi k)}  \approx \frac{4}{3}(Q_1^2+Q_2^2+Q_3^2)+Q_3^2\left(\frac{1}{\sin^2\gamma}-\frac{4}{3}\right)
    \label{20}
\end{align}
Using wave functions of the form
\begin{equation}
\Psi(\beta,\gamma,\theta_i)=F_L(\beta)\eta_K(\gamma)\mathcal{D}_{M,K}^L(\theta_i) \label{21}
\end{equation}
where $\mathcal{D}(\theta_i)$ denote Wigner functions of the Euler angles.
$L$  are the eigenvalues of angular momentum, while  $M$ and $K$ are the eigenvalues of the projections of angular momentum on the laboratory fixed $x$-axis and the body-fixed $x'$-axis respectively. The separation of variables leads to
\begin{equation}
   \Big[ -\frac{1}{\beta^4}\frac{\partial}{\partial\beta} {\beta^4}\frac{\partial}{\partial\beta}+\frac{\bar\Lambda}{\beta^2}+u_1(\beta)\Big]\xi(\beta)=\epsilon \xi(\beta)  \label{22}
\end{equation}
\begin{align}
   \Big[- \frac{1}{\sin3\gamma}\frac{\partial}{\partial\gamma}\sin3\gamma\frac{\partial}{\partial\gamma}+
  \frac{K^2}{4}\Big(\frac{1}{\sin^2\gamma}-\frac{4}{3}\Big) +u_2(\gamma) \Big]\eta_K(\gamma)={\hat\Lambda}\eta_K (\gamma) \label{23}
\end{align}
where $\bar\Lambda=\hat{\Lambda}+L(L+1)/3$ and $\hat\Lambda$ is a parameter coming from the exact separation of the variables, obtained from the $\gamma$ equation. The equation \eqref{23} has been solved \cite{X5} for the potential \eqref{19}  leading to

\begin{equation}
\hat\Lambda= (6c)(n_\gamma+1)-\frac{K^2}{3}  \label{24}
\end{equation}
where $n_\gamma$ is the quantum number related to $\gamma$-oscillations.
\subsection{Common form of the radial part}
We see that Eq .\eqref{eq3} has the same form as  \eqref{22}, obtained in the $\gamma$-unstable case, the only difference being that $\Lambda$ in the first equation is replaced by $\bar\Lambda$ in the axially symmetric prolate deformed nuclei. In what follows we are going to use the symbol $\Lambda$.

In this work, we use the Tietz-Hua potential with a unit depth expressed as \cite{tietz1963potential,hua1990four}
\begin{equation}\label{eq6}
u_1(\beta)=\left[\frac{1-e^{-b_h(\beta-\beta_e)}}{1-c_he^{-b_h(\beta-\beta_e)}}\right]^2
\end{equation}
By inserting the function $f(\beta)=\beta^{-2}\xi(\beta)$ in the radial equation \eqref{eq3}, we obtain:
\begin{equation}\label{eq7}
\left[-\frac{d^2}{d^2\beta}+\frac{\Lambda+2}{\beta^2}+\left[\frac{1-e^{-b_h(\beta-\beta_e)}}{1-c_he^{-b_h(\beta-\beta_e)}}\right]^2\right]f(\beta)=\epsilon f(\beta).
\end{equation}

For a small $\beta$ deformation, the centrifugal potential could be expanded around $\beta=\beta_e$ in a series of powers of $x=(\beta-\beta_e)/\beta_e$ as

\begin{equation}\label{eq8}
V_l(\beta)=\frac{\Lambda+2}{\beta^2}=\frac{\Lambda+2}{\beta_e^2(1+x)^2}
=\frac{\Lambda+2}{\beta_e^2}(1-2x+3x^2-4x^3+...)
\end{equation}
It is sufficient to keep expansion terms only up to the second order. The following form of the potential can be used instead of the centrifugal potential in the Pekeris approximation \cite{pekeris34}

\begin{equation}\label{eq9}
\widetilde{V}_l(\beta)=\frac{\Lambda+2}{\beta_e^2}\left(D_0+D_1\frac{e^{-\alpha x}}{1-c_he^{-\alpha x}}+D_2\frac{e^{-2\alpha x}}{(1-c_he^{-\alpha x})^2}\right)
\end{equation}
where $\alpha=b_h\beta_e$ and $D_i$ is the parameter of coefficients $(i=0,1,2)$. By expanding Eq. \eqref{eq9} up to terms $x^2$, after making some arrangements and combining equal power with Eq. \eqref{eq8}, we obtain the relations between the coefficients and parameters $\alpha$ and $c_h$ as follows:

\begin{eqnarray}\label{eq10}
D_0=1-\frac{1}{\alpha}(1-c_h)(3+c_h)+\frac{3}{\alpha^2}(1-c_h)^2,
      \\ \nonumber
      \lim_{c_h\rightarrow0}D_0=1-\frac{3}{\alpha}+\frac{3}{\alpha^2},
\end{eqnarray}
\begin{eqnarray}\label{eq11}
D_1=\frac{2}{\alpha}(1-c_h)^2(2+c_h)-\frac{6}{\alpha^2}(1-c_h)^3,
      \\ \nonumber
      \lim_{c_h\rightarrow0}D_1=\frac{4}{\alpha}+\frac{6}{\alpha^2},
\end{eqnarray}
\begin{eqnarray}\label{eq12}
D_2=-\frac{1}{\alpha}(1-c_h)^3(1+c_h)+\frac{3}{\alpha^2}(1-c_h)^4,
      \\ \nonumber
      \lim_{c_h\rightarrow0}D_2=-\frac{1}{\alpha}+\frac{3}{\alpha^2},
\end{eqnarray}

Rewriting Eq. \eqref{eq7} by using the new variable $z=c_h e^{-\alpha x}$, we obtain

\begin{align}\label{13}
\Bigg[\frac{d^2}{dz^2}&+\frac{1-z}{z(1-z)}\frac{d}{dz}\nonumber\\+&\left(\frac{\beta_e^2\epsilon}{\alpha^2}(1-z)^2-\frac{\beta_e^2}{\alpha^2}\left(1-\frac{1}{c_h}z\right)^2 -\frac{\Lambda+2}{\alpha^2}\left(D_0+\frac{D_1}{c_h}z(1-z)+\frac{D_2}{c_h^2}z^2\right)\right)\Bigg]f(z)=0
 \end{align}

Using the generalized Nikiforov-Uvarov method \cite{NU}, we obtain the energy spectrum of the $\beta$ part as
\begin{align}\label{eq14}
&\epsilon_{n,\Lambda}=\frac{\Lambda+2}{\beta_e^2}D_0\\ \nonumber -&\frac{\alpha^2}{4\beta_e^2}\left[\frac{\frac{(\Lambda+2)}{\alpha^2c_h^2}(D_2-c_hD_1)+\frac{\beta_e^2}{\alpha^2c_h^2}(1-c_h^2)}
{n+\frac{1}{2}+\sqrt{\frac{\Lambda+2}{\alpha^2c_h^2}D_2+\frac{\beta_e^2}{\alpha^2c_h^2}{(1-c_h)^2}+\frac{1}{4}}}
-\left(n+\frac{1}{2}+\sqrt{\frac{\Lambda+2}{\alpha^2c_h^2}D_2+\frac{\beta_e^2}{\alpha^2c_h^2}{(1-c_h)^2}+\frac{1}{4}}\right)\right]^2
\end{align}
In the following, we obtained the corresponding wave function of the $\beta$ part as

\begin{equation}\label{eq15}
\psi(t)=N_n(1-t)^\mu(1+t)^\nu P_n(2\mu,2\nu-1)(t)
\end{equation}

where $t=1-2z$, $\mu=\sqrt{\frac{\Lambda+2}{\alpha^2}D_0+\frac{\beta_e^2}{\alpha^2}D-\frac{\beta_e^2}{\alpha^2}\epsilon}$ and $\nu=1+\sqrt{\frac{\Lambda+2}{\alpha^2c_h^2}D_2+\frac{\beta_e^2}{\alpha^2c_h^2}D(1-c_h)^2+\frac{1}{4}}$, in which $\Lambda$ for $\gamma$-unstable nuclei it is given by $\Lambda=\tau(\tau+3)$ while for axially symmetric prolate deformed nuclei it should be replaced by $\bar\Lambda=(6c)(n_\gamma+1)+\frac{L(L+1)-K^2}{3}$ . 

$N_n$ is computed via the orthogonality relation of Jacobi polynomials:
\begin{equation}\label{eq16}
N_n=\left(\frac{\nu+n}{2\mu(\mu+\nu+n)}\right)^{-\frac{1}{2}}
\left(\frac{\left(\Gamma(2\mu+1)\Gamma(n+1)\right)^2}{\Gamma(2\mu+n+1)}\frac{\Gamma(2\nu+n)}{n!\Gamma(2\nu+2\mu+n)}\right)^{-\frac{1}{2}}
\end{equation}
Note that to compute this constant we required the following parameterization $c_h=e^{-\alpha}$.

\section{$B(E2)$ transition rates}\label{S2}

The $B(E2)$ transition rates from an initial to a final state are given by \cite{edmonds}
      \begin{align}
B(E2;s_i,L_i  \rightarrow s_f,L_f)  =\frac{5}{16\pi} \frac{\mid \left<s_f,L_f\mid\mid T^{(E2)} \mid\mid s_i,L_i\right>\mid^2}{(2L_i+1)}\nonumber\\
    =\frac{2L_f+1}{2L_i+1}B(E2;s_f,L_f  \rightarrow s_i,L_i)
  \label{3.1}
\end{align}
where $s$ denotes quantum numbers other than the angular momentum $L$.
\subsection{$B(E2)$s for $\gamma$-unstable nuclei}
  In the general case the quadrupole operator is defined as \cite{wilet}
 \begin{align}
      T_{M}^{(E2)}=t\alpha_2&=t\beta\left[\mathcal{D}^{(2)}_{M,0}(\theta_i)\cos\gamma  +\frac{1}{\sqrt{2}}\Big( \mathcal{D}^{(2)}_{M,2}(\theta_i)
      +\mathcal{D}^{(2)}_{M,-2}(\theta_i) \Big)\sin\gamma \right]
  \label{3.2}
\end{align}
 where $\mathcal{D}(\theta_i)$ denotes the Wigner functions of Euler angles and $t$ is a scale factor.
Then, the full symmetrized wave function is given by
\begin{equation}
\Psi(\beta,\gamma,\theta_i)=\beta^{-2}\chi_{n,\tau}(\beta)\Phi_{\tau}(\gamma,\theta_i)
 \label{3.3}
\end{equation}
The radial function $\chi(\beta)$ is given by Eq. \eqref{eq15}, while the angular functions $\Phi_{\tau}(\gamma,\theta_i)$ have the form \cite{bes}
\begin{equation}
\Phi_{\tau}(\gamma,\theta_i)=\frac{1}{4\pi}\sqrt{\frac{(2\tau+3)!!}{\tau!}}\left( \frac{\alpha_2}{\beta^2}\right)^{\tau}
 \label{3.4}
\end{equation}
where $\alpha_2$ is defined in Eq. \eqref{3.2}. From Eqs. \eqref{3.1} and \eqref{3.4} one obtains \cite{bonat04}
\begin{equation}
B(E2;L_{n,\tau}\rightarrow(L+2)_{n',\tau+1})=\frac{(\tau+1)(4\tau+5)}{(2\tau+5)(4\tau+1)}t^2I_{n',\tau+1;n,\tau}^2
\label{3.5}
\end{equation}
with
\begin{equation}
I_{n',\tau+1;n,\tau}= \int_0^{\infty} \beta \xi_{n',\tau+1}(\beta)\xi_{n,\tau}(\beta)\beta^4d\beta
\label{3.6}
\end{equation}
The $\tau$ dependence of the radial wave functions $R_n(\beta)$ of Eq. \eqref{eq15} is contained in $\mu$ and $\nu$, which in turn to  contain $\Lambda=\tau(\tau+3)$
\subsection{$B(E2)$s for axially symmetric prolate deformed nuclei}
The $B(E2)$ transition rates for axially deformed nuclei around $\gamma=0$ read \cite{bijker03}
    \begin{align}
B(E2;nLn_{\gamma}K\longrightarrow n'L'n'_{\gamma}K')
=\frac{5}{16\pi}t^2\langle L,K,2,K'-K|L',K'\rangle^2 I^2_{n,L;n',L'}C^2_{n_{\gamma},K;n'_{\gamma},K'}
\label{3.7}
  \end{align}
where 
\begin{align}
I_{n,L;n',L'}=\int_0^{\infty} \beta \xi_{n',L'}(\beta)\xi_{n,L}(\beta)\beta^4d\beta 
\label{3.8}
  \end{align}
is the integral over $\beta$,  while $C_{_{n_{\gamma}K,n'_{\gamma}K'}}$ contains the integral over $\gamma$. For $\Delta K=0$ corresponding to transitions  ($g.s.\rightarrow g.s., \gamma\rightarrow\gamma, \beta\rightarrow\beta$ and $\beta\rightarrow g.s. $), the $\gamma-$integral part reduces to the orthonormality condition of the $\gamma$-wave functions : $C_{_{n_{\gamma}K,n'_{\gamma}K'}}=\delta_{_{n_{\gamma},n'_{\gamma}}}\delta_{_{K,K'}}$. While for $\Delta K=2$ corresponding to transitions ($\gamma\rightarrow g.s., \gamma\rightarrow\beta$), this
integral takes the form.
\begin{align}
C_{n_{\gamma}K,n'_{\gamma}K'}=\int \sin\gamma\eta_{n_{\gamma}K}\eta_{n'_{\gamma}K'}|\sin3\gamma|d\gamma
\label{3.9}
  \end{align}
In the next sections, all values of $B(E2)$ are calculated in units of $B(E2;2+_{1}\longrightarrow 0^+_{1})$.
\section{results}
The theoretical predictions for the energy spectra of the $g.s.$, $\beta$ and $\gamma$ bands for $\gamma$-unstable and axially symmetric deformed nuclei are obtained from equation \eqref{eq14} by fitting the model parameters on the experimental data using the quality measure : 
\begin{equation}\label{3.10}
\sigma=\sqrt{\frac{\sum_{i=1}^N(E_i(Exp)-E_i(th))^2}{(N-1)E(2_1^+)}}
\end{equation}
where $N$ denotes the number of the states, while $E_i(exp)$ and $E_i(th)$ represent the experimental and theoretical energies of the $i^{th}$ level, respectively. $E(2^+_i)$ is the energy of the first excited level of the $g.s.$ band.

In Table 1 and Table 2, we present the obtained numerical results for the $g.s.$ bandhead $R_{4/2}=E(4_g^+)/E(2_g^+)$ ratios as well as those of the $\beta$ and $\gamma$ bandheads, normalized to the $2^+_g$ level, namely : $R_{0/2}=E(0_{\beta}^+)/E(2_g^+)$ and $R_{2/2}=E(2_{\gamma}^+)/E(2_g^+)$ respectively for $\gamma$-unstable (Table 1) and axially symmetric prolate nuclei (Table 2).

The calculations have been carried out for several nuclei from $^{98}Ru$ to  $^{200}Pt$ in the $\gamma$-unstable case and from $^{150}Na$ to $^{250}Cf$ in the axially symmetric prolate one.

From Table \eqref{table1}, one ca see that over 54 studied $\gamma$-unstable nuclei, 87\% are well reproduced with $\sigma <1$ in respect to the experimental data. While from Table \eqref{table2}, one can observe that only 50 \% of the 59 studied axially symmetric nuclei have $\sigma<1$. This is due to the observed discrepancy in the $\beta$-band in respect to the experiment. So, at this stage, the Tietz-Hua potential seems to be appropriate for predictions of energy spectra of $\gamma$-unstable nuclei. But, its ability to reproduce well the experimental spectra for all kind nuclei could be improved within the deformation dependent mass formalism (DDMF) \cite{bonatsos2010bohr} as it has been done with Davidson potential in Ref. \cite{bonat11}. The DDMF is well known to have an effect on the enhancement of the calculations precision of the energy levels of nuclei \cite{bonat13,bonat11,CLO15P}.

In Figures \eqref{fig1}-\eqref{fig2}, are presented the spectra of $^{162}$Yb and $^{166}$Er compared with the experimental levels \cite{data}. From these figures, we can see the overall agreement between our results  and the experiment. In these figures, we also show some transition rates which are very well reproduced in comparison with the experimental data. 

\begingroup
\begin{figure*}
\begin{center}
\includegraphics[scale=.85]{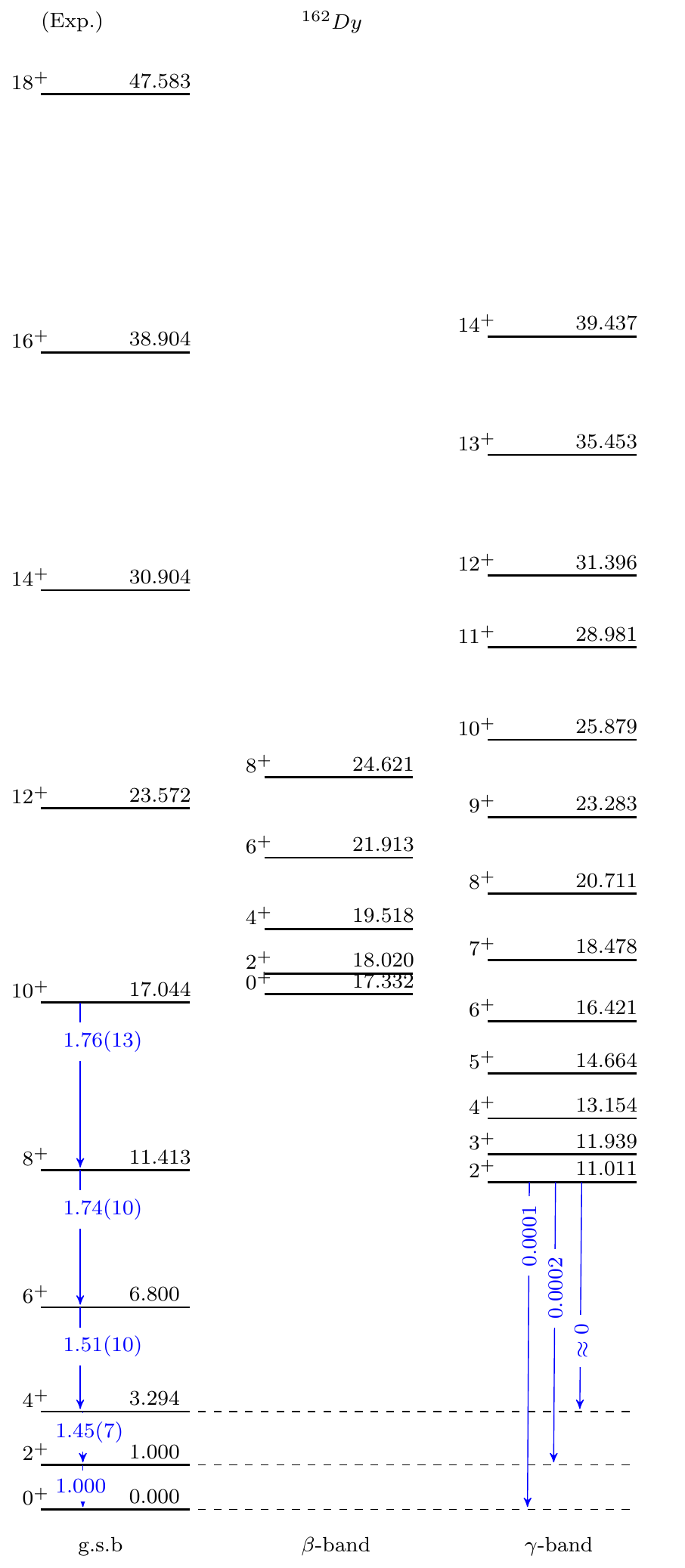} 
\includegraphics[scale=.85]{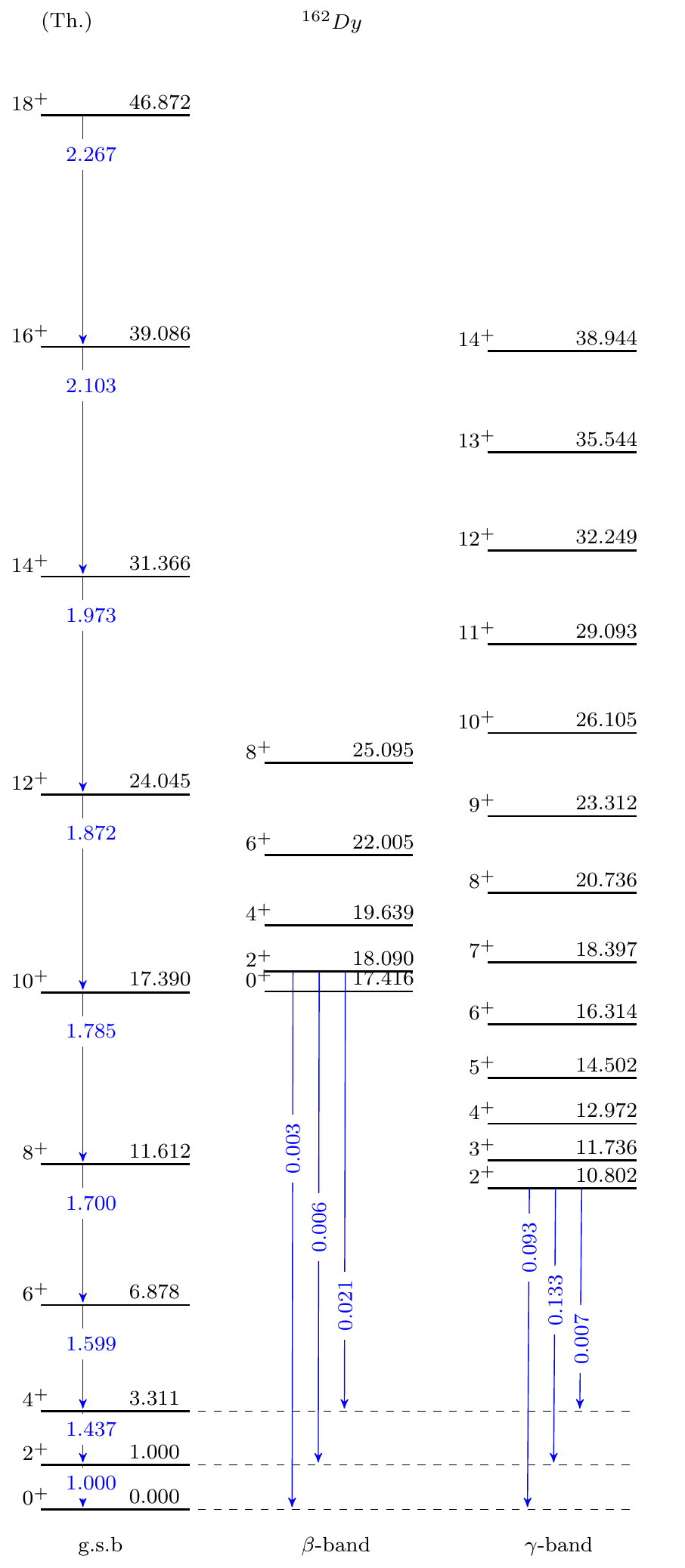}
  \caption{ The theoretical energy spectra and some $B(E2)$ transitions for the ground (g.s.), $\gamma$ and $\beta$ bands, are compared with the experimental data \cite{data} for $^{162}$Dy.     }\label{fig1}
\end{center}
\end{figure*}
\begin{figure*}
\begin{center}
\includegraphics[scale=.85]{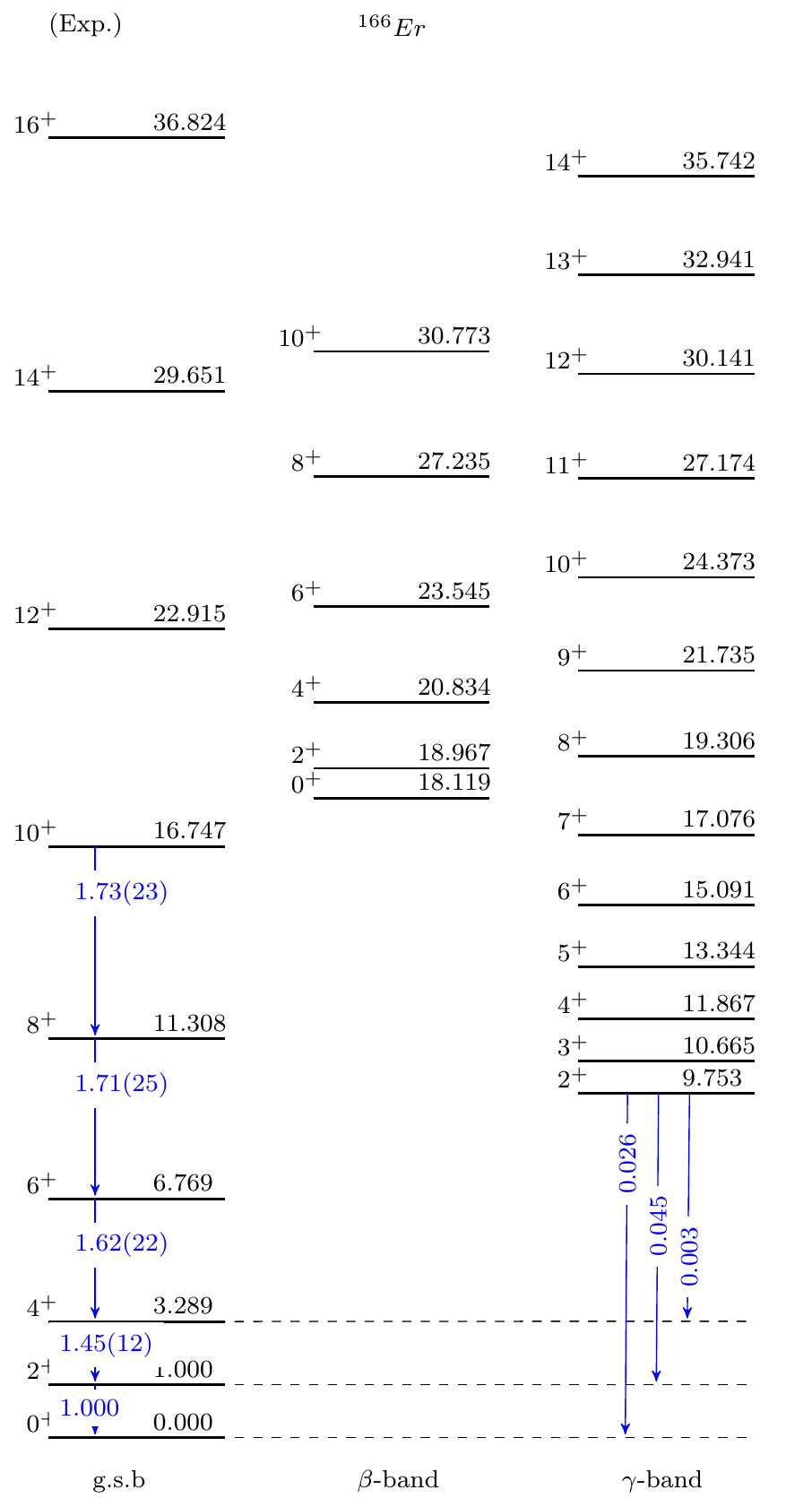} 
\includegraphics[scale=.85]{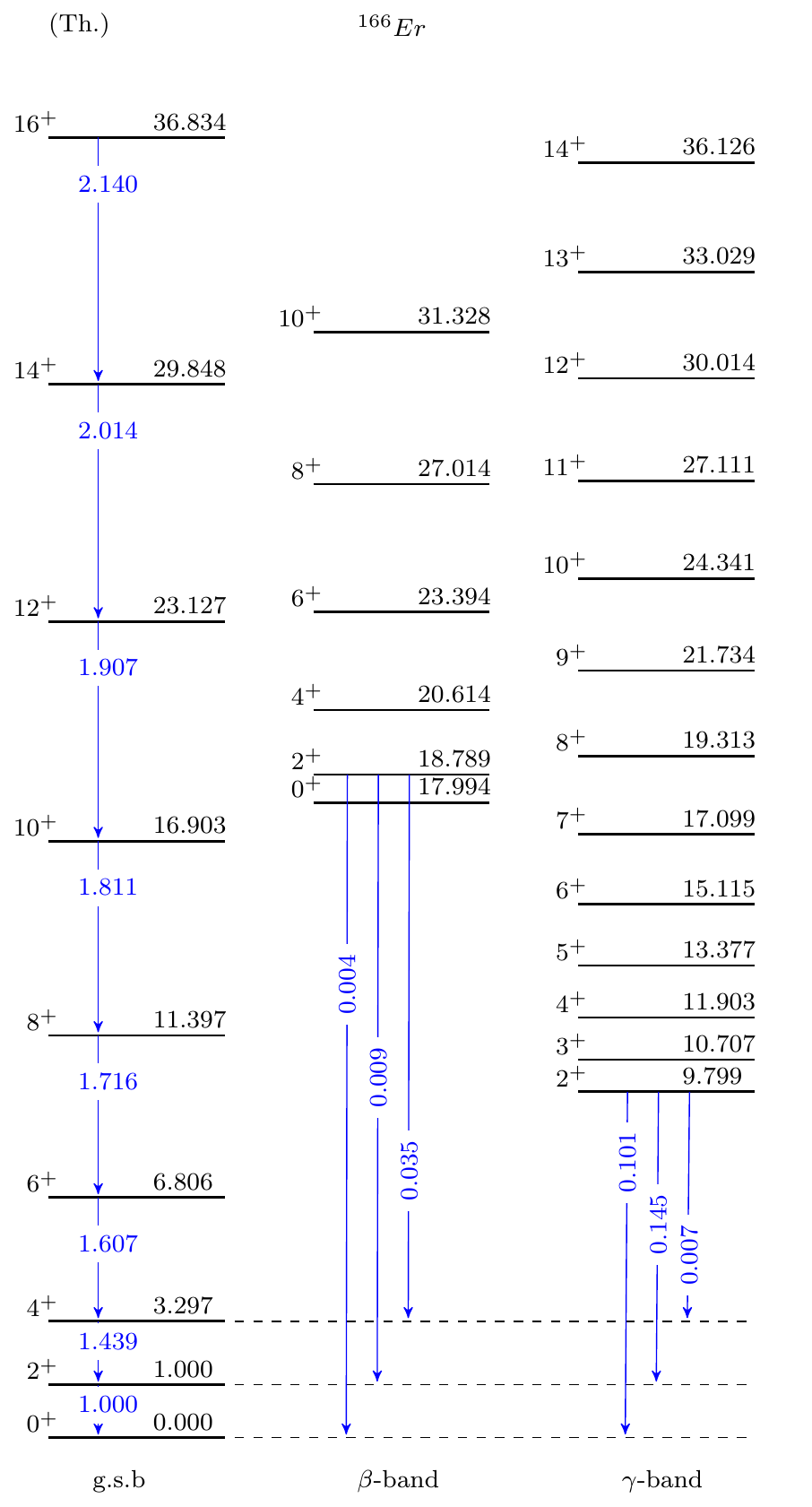}
  \caption{ The theoretical energy spectra and some $B(E2)$ transitions for the ground (g.s.), $\gamma$ and $\beta$ bands, are compared with the experimental data \cite{data} for $^{166}$Er.      }\label{fig2}
\end{center}
\end{figure*}
\endgroup
The exhaustive results for transition rates of ground-ground, $\beta$-ground and $\gamma$-ground transitions are given in Table \eqref{table3} for several $\gamma$-unstable nuclei and  in Table \eqref{table4} for several axially symmetric prolate ones. Also, in  both tables, are listed the obtained results by Manning-Rosen $V_{MR}(\beta)$ \cite{chabab2016electric},  Kratzer $V_K(\beta)$ \cite{bonat13}, Davidson $V_D(\beta)$ \cite{bonat11} and Morse $V_M(\beta)$ \cite{Inci11} potentials for comparison.

\par Statistics for such a comparison are presented in Figure \eqref{fig4} where one can clearly see that over the 34 studied $\gamma$-unstable nuclei and 38 axially symmetric prolate ones, the larger part of these nuclei is well reproduced with $\sigma <1$ by Tietz-Hua potential (large Pie charts). In the second rank comes Manning-Rosen potential \cite{chabab2016electric} which should be naturally followed by Morse \cite{Inci11}  as can be seen in the comparison between potentials without including the Tietz-Hua one (small Pie charts). Such a fact  has already been proved in \cite{chabab2016electric}. But, when comparing all potentials including Tietz-Hua, we have to notice that this latter appropriated the majority of nuclei which were well reproduced by Morse as can be seen from Tables \eqref{table3}-\eqref{table4}. While Davidson \cite{bonat11} and Kratzer \cite{bonat13} potentials occupy the last places respectively.

\begingroup
\begin{figure*}
\begin{center}
	\includegraphics[width=0.49\textwidth]{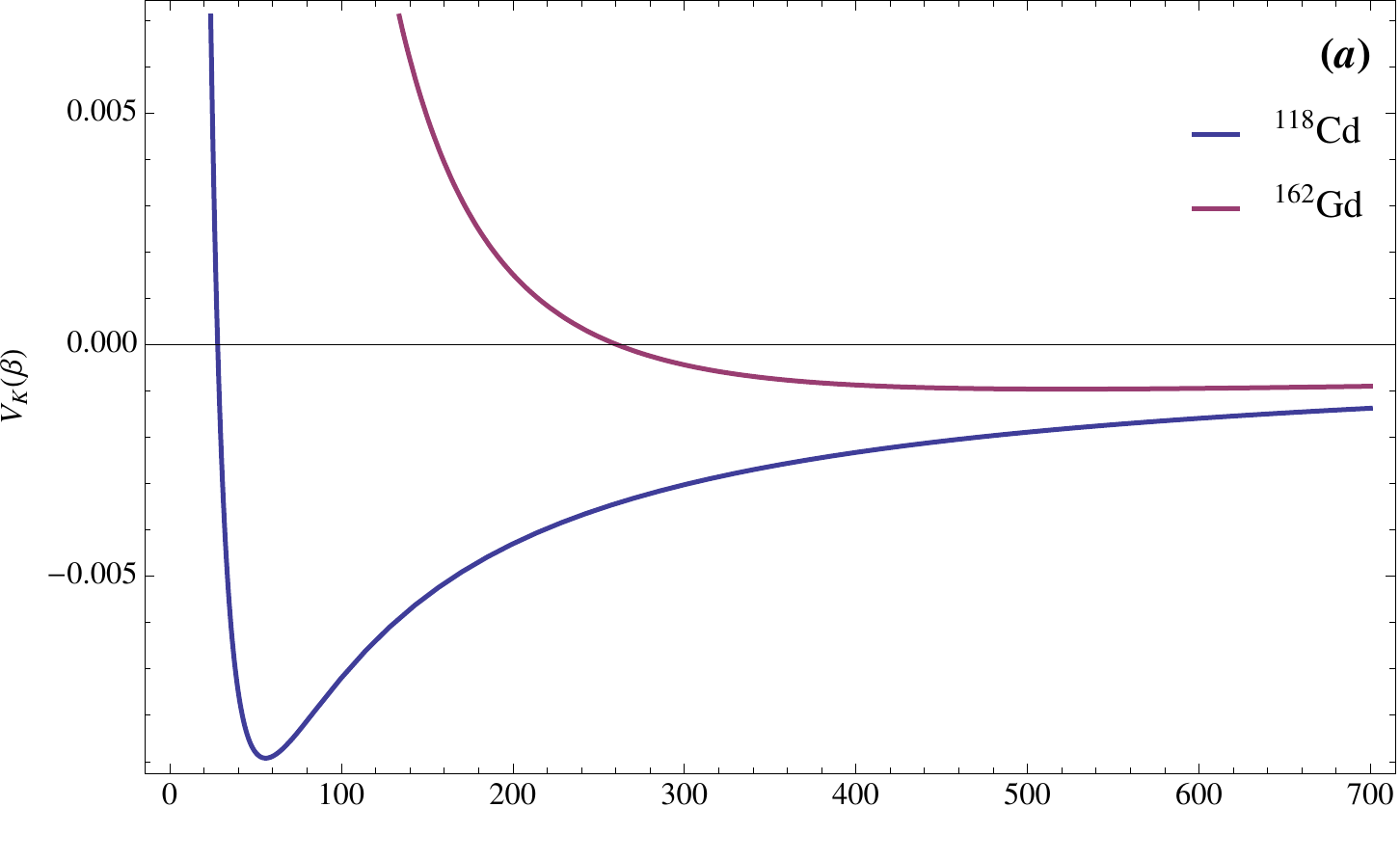}
\hspace{0.2cm}
	\includegraphics[width=0.47\textwidth]{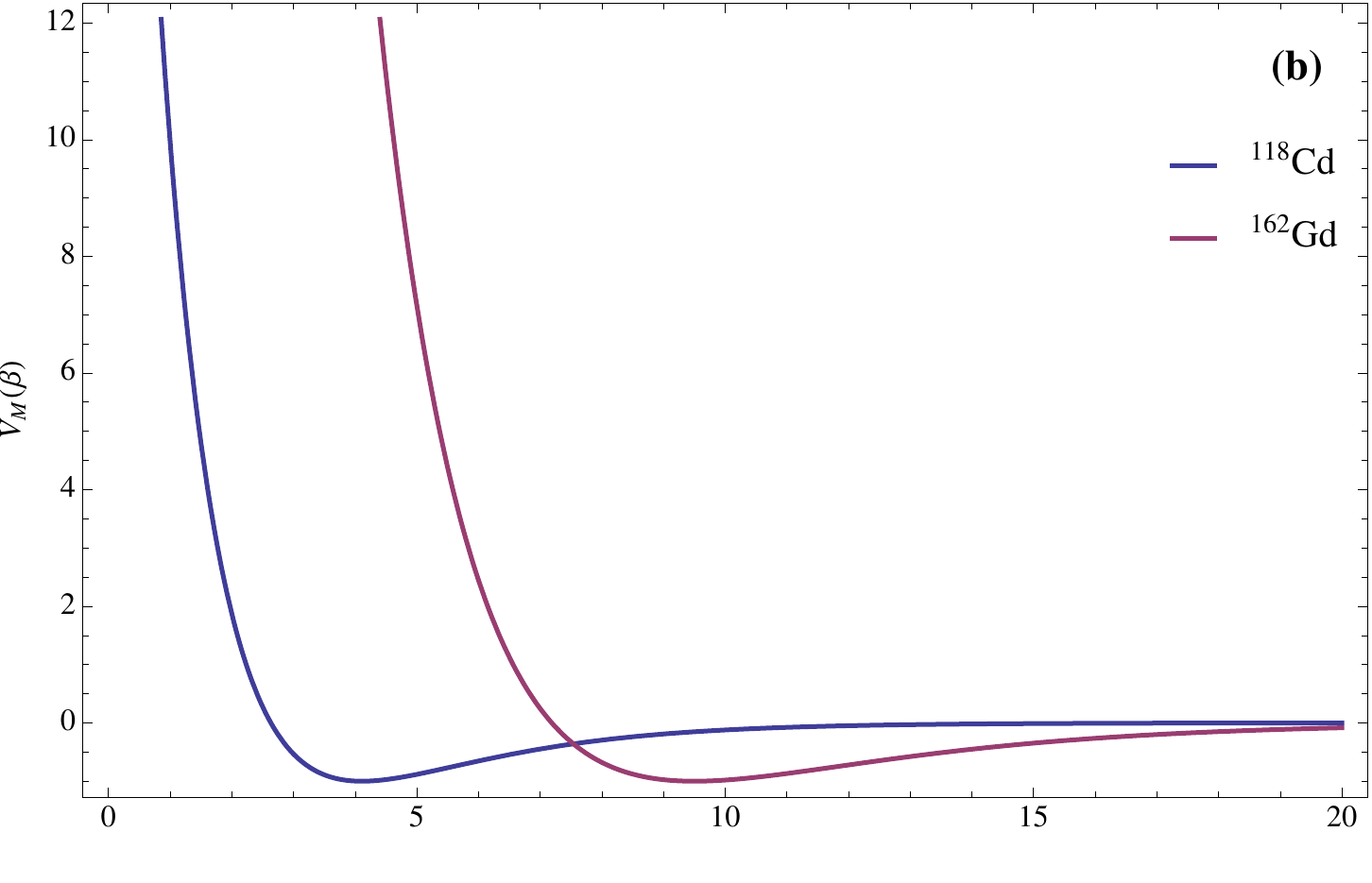}
\includegraphics[width=0.49\textwidth]{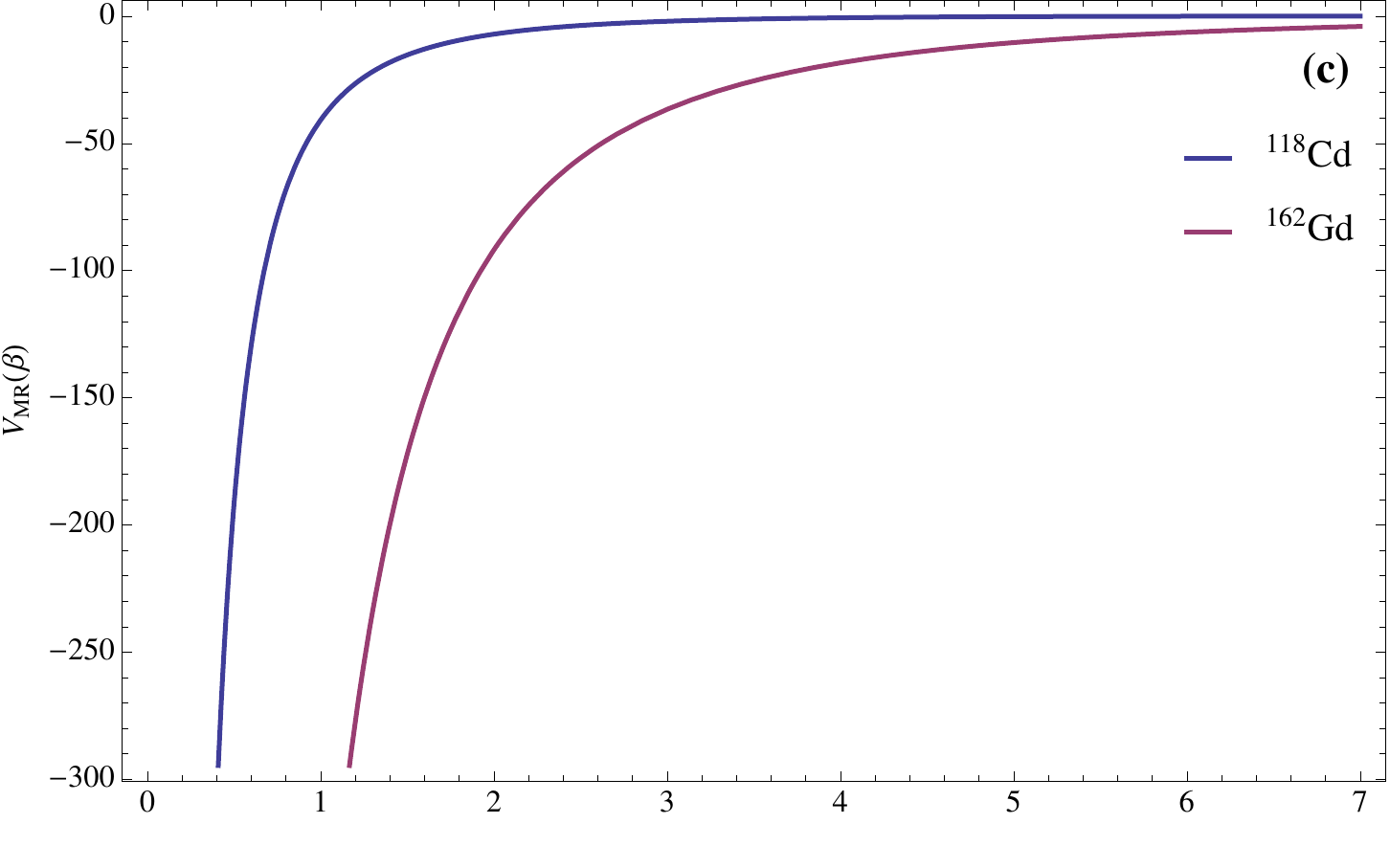}
\hspace{0.2cm}
	\includegraphics[width=0.47\textwidth]{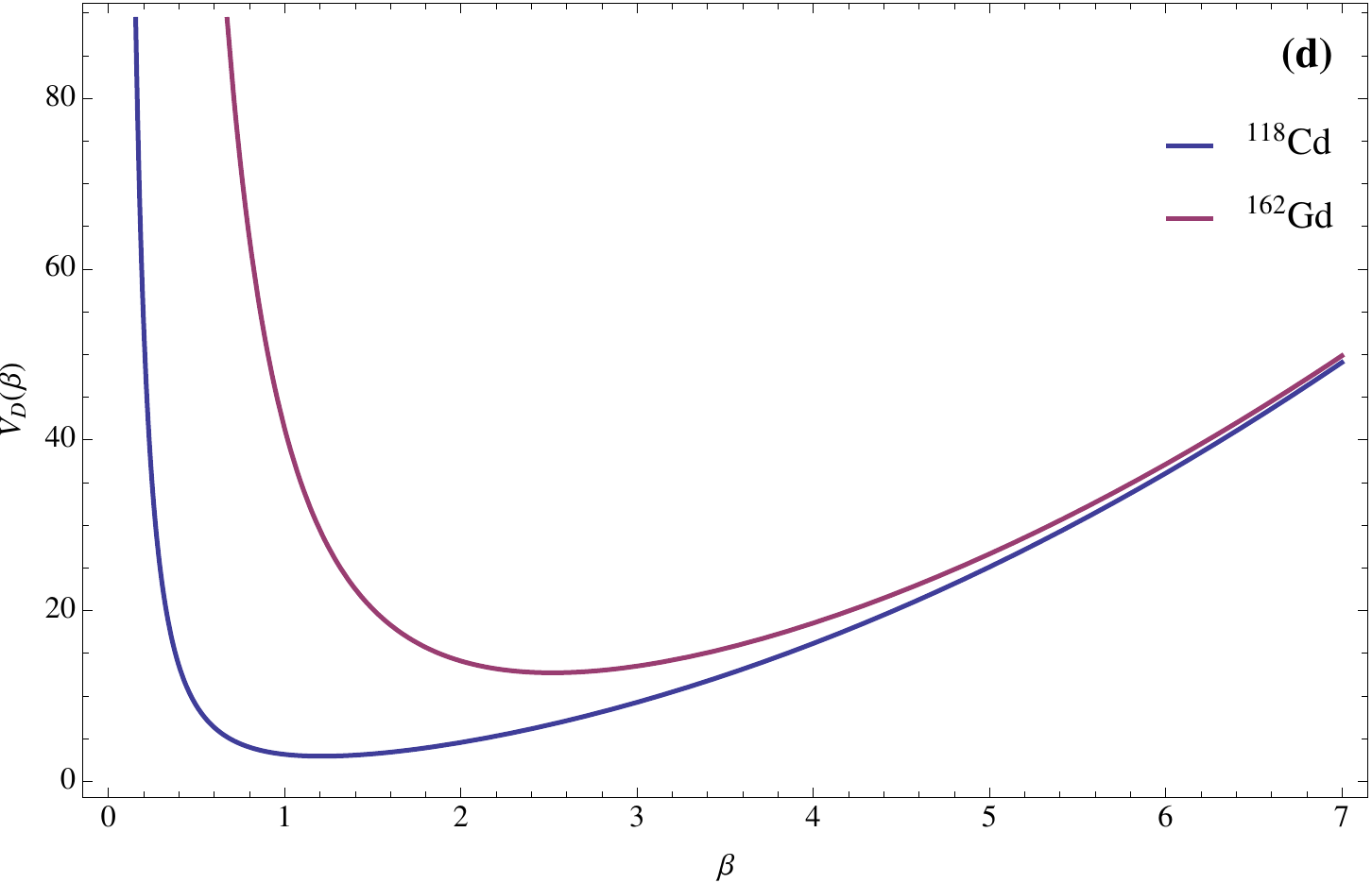}
\includegraphics[width=0.49\textwidth,left]{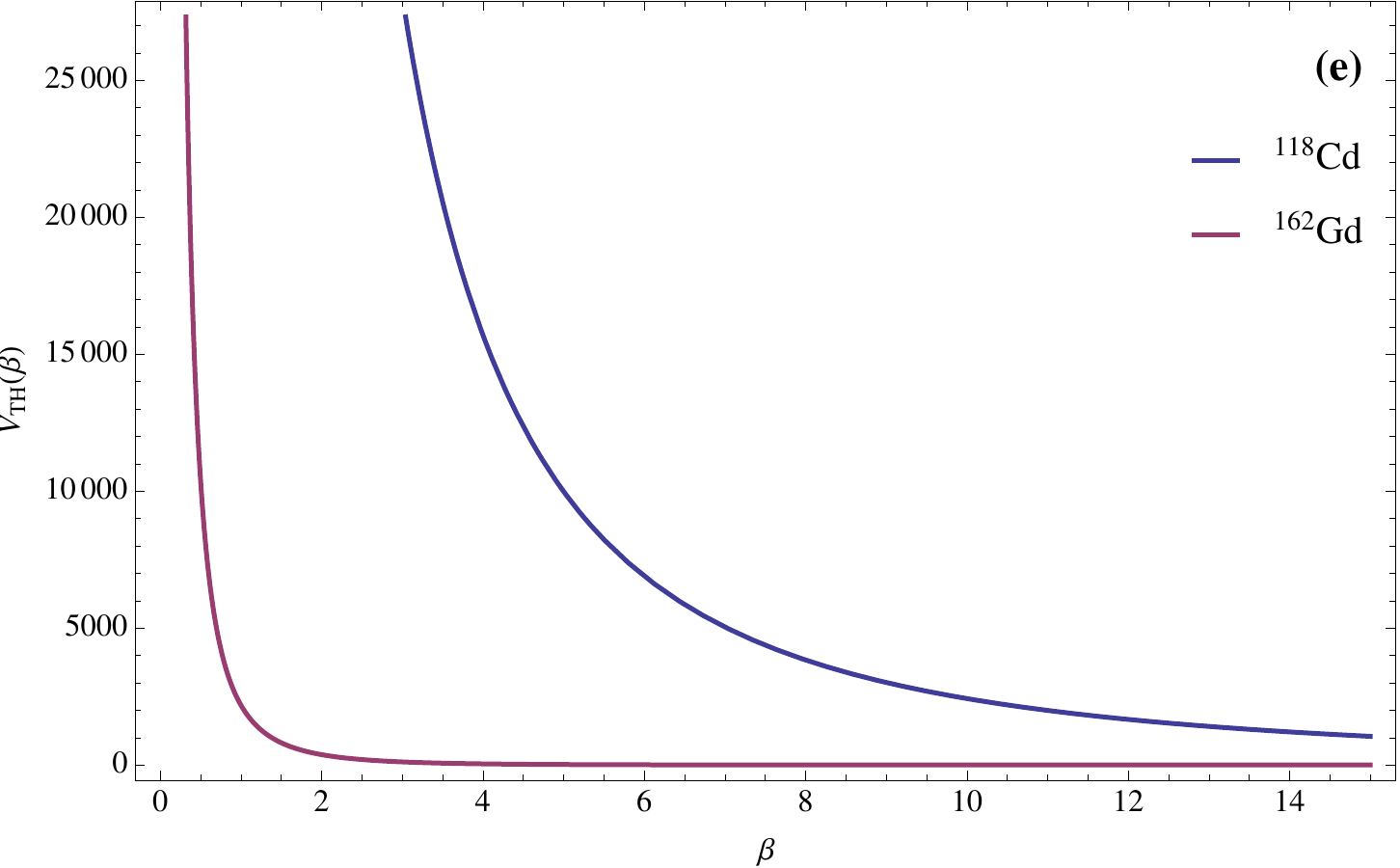}
  \caption{ (Color online) Evolution of   Kratzer $V_K(\beta)$ (a), Morse $V_M(\beta)$ (b), Manning-Rosen $V_{MR}(\beta)$ (c),  Davidson $V_D(\beta)$ (d) and  Tietz-Hua $V_{TH}(\beta)$ (e) potentials, for $^{118}Cd$ and $^{162}Gd$ nuclei. The quantities shown are dimensionless. }\label{fig3}
   \end{center}
\end{figure*}
\endgroup

The efficience of Tietz-Hua  potential in reproducing the experimental data for transitions rates in comparison with the other model potentials is due to its shape particularly for $\beta$ values beyond the potential minimum. Indeed, it has already been shown \cite{chabab2016electric} that while the considered potential is flatter beyond its minimum, the precision of calculated rates is greater. So, from Figure \eqref{fig3}, one can see that the Tietz-Hua potential is the best candidate for this purpose.  
   
\begingroup
\begin{figure*}
\begin{center}
	\includegraphics[width=0.49\textwidth]{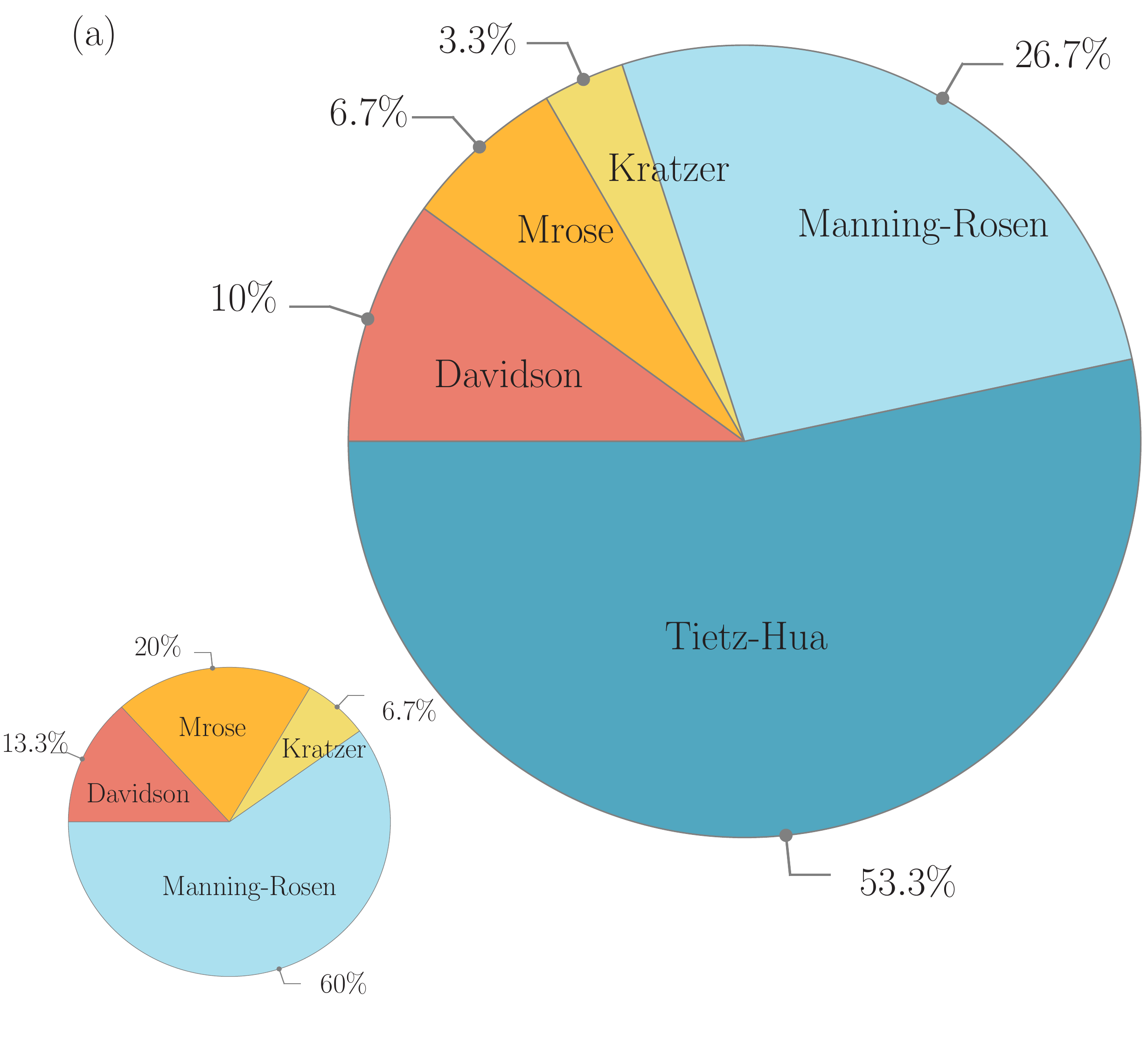}
	\includegraphics[width=0.49\textwidth]{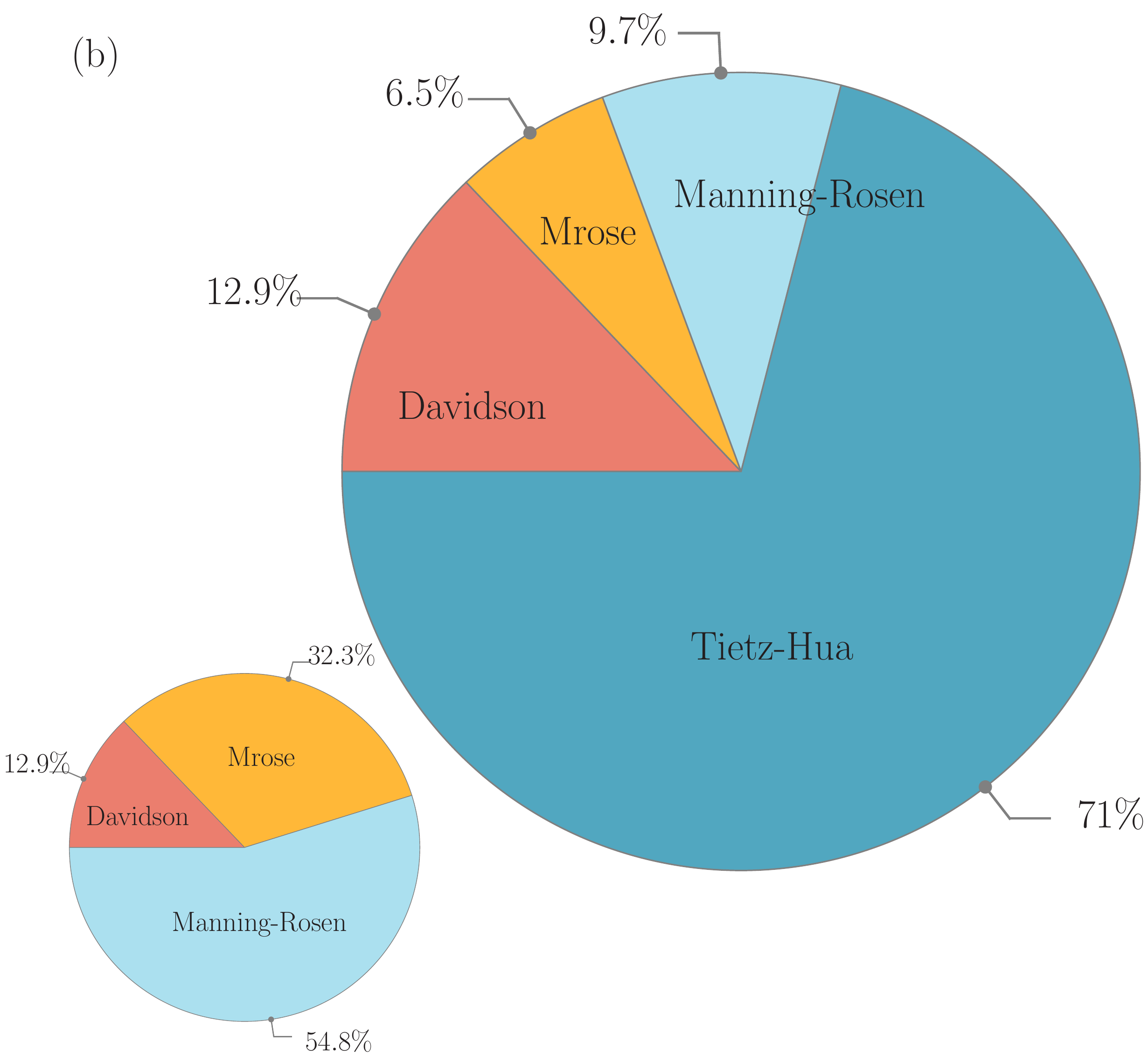}
 \caption{ (Color online) The percentage of the best value of the $rms$  in Tables (\ref{table3}-\ref{table4}) obtained by Tietz-Hua $V_{TH}(\beta)$,  Manning-Rosen $V_{MR}(\beta)$ \cite{chabab2016electric},  Kratzer $V_K(\beta)$ \cite{bonat13}, Davidson $V_D(\beta)$ \cite{bonat11} and Morse $V_M(\beta)$ \cite{Inci11} potentials, in the $\gamma$-unstable case (left) and the rotational case (right). The small Pie charts correspond to the statistical study without  considering the Tietz-Hua potential. }\label{fig4}
\end{center}
\end{figure*}
\endgroup

\par Another characteristic, for axially symmetric prolate deformed nuclei, which has to be cheked, is obviously the  energy level degeneracy  appearing between the $\beta$ and $\gamma$ bandheads, namely $0^+_{\beta}$ and $2^+_{\gamma}$ states.  This property has  been first suggested in the work of  Alhassid-Whelan arc of regularity \cite{alhassid1991chaotic}, such an arc moving inside the symmetry triangle of the IBA model \cite{iachello1987IBM},  connecting the U(5) and SU(3) vertices. An experimental confirmation was realized \cite{jolie2004experimental} to identify the atomic nuclei located close to the regular region of the Casten triangle \cite{casten2000nuclear,Casten:2006} noted by Alhassid and Whelan. Among the studied nuclei  in Table \eqref{table2}, the possible candidates satisfying the theoretical signature $|E(2_{\gamma}^+)-E(0^+_{\beta})|/E(2^+_{\gamma})\leq 0.05$ \cite{alhassid1991chaotic} and the experimental one $|E(2_{\gamma}^+)-E(0^+_{\beta})|/E(2^+_{\gamma})\leq 0.025$ \cite{jolie2004experimental} are $^{158}Gd$, $^{178}Hf$ and $^{250}Cf$. Comparing these results with those of \cite{bonat7}, one can observe that our calculations with Tietz-Hua potential do not predict the additional two nuclei in \cite{bonat7}, namely:$^{158}Dy$ and $^{236}U$. This is due to the above cited discrepancies of the present model. However, as it was also mentioned above, one can remedy to this problem by reconsidering the present study within DDMF \cite{bonat13,bonat11,CLO15P}. But, here we have to notice that the nucleus $^{170}Er$ which has been already predicted in \cite{bonat7} as belonging to Alhassid-Whelan arc of regularity is inconsistent with the experimental results, while our predictions for such a nucleus  are coherent with these latter.

\section{Conclusion}
In this work, new solutions for the Bohr Hamiltonian are obtained with Tietz-Hua potential, here used as a recall potential for $\beta$-vibration and a harmonic oscillator one for $\gamma$-vibration. The calculated energy spectra for several $\gamma$-unstable and axially symmetric prolate nuclei, within the present model, are satisfactory in comparison with the experimental data. But, as it has been mentioned above, such calculations could be improved in the framework of the deformation dependent mass formalism as it has been proved with Davidson potential in Ref \cite{bonatsos2010bohr}. As to the transition probabilities, it has been shown that Tietz-Hua potential presents an absolute prevalence when reproducing the experimental results. Such a fact is due to the flatness of this potential for large values of the $\beta$ variable in comparison with other model potentials.    

\newpage
\begin{longtable}{lllllllllllllll}
\caption{ The comparison of theoretical predictions of the Bohr Hamiltonian with  Tietz-Hua potential  to experimental data for the ground state bandhead $R_{4/2}=E(4^+_{g})/E(2^+_{g})$ ratios, as well as those of the $\beta$ and $\gamma$ bandheads, normalized to the $2^+_{g}$ state and labelled by $R_{0/2}=E(0^+_\beta)/E(2^+_{g})$ and $R_{2/2}=E(2^+_\gamma)/E(2^+_{g})$ respectively. $L_g$, $L_{\beta}$ and $L_{\gamma}$ characterized the angular momenta of the highest levels of the ground state, $\beta$ and $\gamma$ bands respectively, included in the fit.\label{table1}} \\
\hline
nuleus & $R_{4/2}$ & $R_{4/2}$ &$R_{0/2}$ & $R_{0/2}$ & $R_{2/2}$&$R_{2/2}$&$\beta_{e}$&$b_h$&$L_g$&$L_\beta$&$L_\gamma$&$n$&$\sigma$\\
&exp&th&exp&th&exp&th\\
\hline
\endfirsthead
\caption{(continued)}\\
\hline 
nuleus & $R_{4/2}$ & $R_{4/2}$ &$R_{0/2}$ & $R_{0/2}$ & $R_{2/2}$&$R_{2/2}$&$\beta_{e}$&$b_h$&$L_g$&$L_\beta$&$L_\gamma$&$n$&$\sigma$\\
&exp&th&exp&th&exp&th\\
\hline
\endhead
\hline
\endfoot
$ ^{  98}$Ru & $2.142$ & $2.334$&$ 2.026$ & $3.951$&$2.168$ & $2.334$&$8.6$&$0.001$&$24$&$0$&$4$&$15$&$0.965$\\

$ ^{100}$Ru & $2.273$ & $2.411$&$ 2.095$& $5.671$&$2.525$ & $2.411$&$12.2$&$0.001$&$28$&$0$&$4$&$17$&$1.584$\\

$ ^{102}$Ru & $2.329$ & $2.338$&$ 1.986$ & $4.016$&$2.322$ & $2.338$&$8.7$&$0.002$&$16$&$0$&$5$&$12$&$0.742$\\

$ ^{104}$Ru & $2.482$ & $2.408$&$ 2.761$ & $3.235$&$2.495$ & $2.408$&$4.8$&$0.621$&$8$&$2$&$8$&$12$&$0.472$\\

$ ^{102}$Pd & $2.293$ & $2.419$&$ 2.863$ & $5.995$&$2.758$ & $2.419$&$12.2$&$0.01$&$26$&$4$&$4$&$18$&$1.954$\\

$ ^{104}$Pd & $2.381$ & $2.338$&$ 2.399$ & $4016$&$2.414$ & $2.338$&$8.7$&$0.002$&$18$&$2$&$4$&$13$&$0.750$\\

$ ^{106}$Pd & $2.402$ & $2.310$&$ 2.215$ & $3.641$&$2.204$ & $2.310$&$7.9$&$0.002$&$16$&$4$&$5$&$14$&$0.808$\\

$ ^{108}$Pd & $2.416$ & $2.426$&$ 2.426$ & $2.990$&$2.146$ & $2.426$&$4.9$&$0.695$&$14$&$4$&$4$&$12$&$0.500$\\
\hline
\end{longtable}
\newpage

\begin{longtable}{llllllllllllllll}
\caption{ The comparison of theoretical predictions of the axially symmetric prolate deformed nuclei of the Bohr Hamiltonian with  Tietz-Hua potential to experimental data (nuclei with $R_{4/2} > 2.9$) for the ground state bandhead $R_{4/2}=E(4^+_{g.s.})/E(2^+_{g.s.})$ ratios, as well as those of the $\beta$ and $\gamma$ bandheads, normalized to the $2^+_{g.s.}$ state and labelled by $R_{0/2}=E(0^+_\beta)/E(2^+_{g.s.})$ and $R_{2/2}=E(2^+_\gamma)/E(2^+_{g.s.})$ respectively. $L_g$, $L_{\beta}$ and $L_{\gamma}$ characterized the angular momenta of the highest levels of the ground state, $\beta$ and $\gamma$ bands respectively, included in the fit.\label{table2}} \\
\hline
nuleus & $R_{4/2}$ & $R_{4/2}$ &$R_{0/2}$ & $R_{0/2}$ & $R_{2/2}$&$R_{2/2}$&$\beta_{e}$&$b_h$&c&$L_g$&$L_\beta$&$L_\gamma$&$n$&$\sigma$\\
&exp&th&exp&th&exp&th\\
\hline
\endfirsthead
\caption{(continued)}\\
\hline 
nuleus & $R_{4/2}$ & $R_{4/2}$ &$R_{0/2}$ & $R_{0/2}$ & $R_{2/2}$&$R_{2/2}$&$\beta_{e}$&$b_h$&c&$L_g$&$L_\beta$&$L_\gamma$&$n$&$\sigma$\\
&exp&th&exp&th&exp&th\\
\hline
\endhead
\hline
\endfoot
$ ^{150}$Nd & $2.93$ & $3.14$&$ 5.19$ & $7.36$&$8.16$ & $8.60$&$6.8$&$0.003$&3.5&$14$&$6$&$4$&$13$&$1.083$\\

$ ^{152}$Sm & $3.01$ & $3.18$&$ 5.62$ & $8.38$&$8.92$ & $10.52$&$7.8$&$0.001$&$4.3$&$16$&$14$&$9$&$23$&$1.716$\\

$ ^{154}$Sm & $3.25$ & $3.30$&$ 13.41$ & $14.01$&$17.57$ & $18.59$&$8.2$&$0.351$&$6.7$&$16$&$6$&$7$&$17$&$0.601$\\

$ ^{154}$Gd & $3.01$ & $3.24$&$ 5.53$ & $11.15$&$8.10$ & $8.45$&$11.4$&$0.001$&$3.0$&$26$&$26$&$7$&$32$&$4.230$\\

$ ^{156}$Gd & $3.24$ & $3.28$&$ 11.79$ & $16.15$&$12.97$ & $14.29$&$16.3$&$0.001$&$5.1$&26 &12 &16 &34&$2.639$\\

$ ^{158}$Gd & $3.29$ & $3.30$&$ 15.05$ & $15.48$&$14.93$ & $15.16$&$8.4$&$0.330$&$5.3$&$12$&$6$&$6$&$14$&$0.242$\\

$ ^{160}$Gd & $3.30$ & $3.31$&$ 17.62$ & $17.86$&$13.13$ & $13.08$&$8.6$&$0.432$&$4.4$&$16$&$4$&$8$&$17$&$0.172$\\

$ ^{162}$Gd & $3.29$ & $3.31$&$ 19.93$ & $19.91$&$11.98$ & $11.90$ & $9.5$ & $0.294$ &$4.0$&$14$&$0$&$4$&$10$&$0.096$\\
\hline
\end{longtable}

\newpage
\footnotesize{
\begin{longtable}{lllllllllllllll}
\caption{
Comparison of experimental data (Exp) \cite{data}  for several B(E2) ratios of  $\gamma$-unstable nuclei to predictions by the Bohr Hamiltonian with the Tietz-Hua potential (TH), Manning-Rosen potential (MR) \cite{chabab2016electric}, Morse potential (M) \cite{Inci11}, Davidson potential (D) \cite{bonat11} and Kratzer potential (K) \cite{bonat13}.  \label{table3}}\\
\hline
nucl.&& $\frac{4_g\rightarrow 2_g}{2_g\rightarrow 0_g}$ & $\frac{6_g\rightarrow 4_g}{2_g\rightarrow 0_g}$ &$\frac{8_g\rightarrow 6_g}{2_g\rightarrow 0_g}$ & $\frac{10_g\rightarrow 8_g}{2_g\rightarrow 0_g}$ & $\frac{2_{\gamma}\rightarrow 2_g}{2_g\rightarrow 0_g}$&$\frac{2_{\gamma}\rightarrow 0_g}{2_g\rightarrow 0_g}$&$\frac{0_{\beta}\rightarrow 2_g}{2_g\rightarrow 0_g}$&$\frac{2_{\beta}\rightarrow 0_g}{2_g\rightarrow 0_g}$&rms\\
&&&&&&&$\times 10^3$&&$\times 10^3$  \\
\hline
\endfirsthead
\caption{(continued)}\\
\hline
nucl.&& $\frac{4_g\rightarrow 2_g}{2_g\rightarrow 0_g}$ & $\frac{6_g\rightarrow 4_g}{2_g\rightarrow 0_g}$ &$\frac{8_g\rightarrow 6_g}{2_g\rightarrow 0_g}$ & $\frac{10_g\rightarrow 8_g}{2_g\rightarrow 0_g}$ & $\frac{2_{\gamma}\rightarrow 2_g}{2_g\rightarrow 0_g}$&$\frac{2_{\gamma}\rightarrow 0_g}{2_g\rightarrow 0_g}$&$\frac{0_{\beta}\rightarrow 2_g}{2_g\rightarrow 0_g}$&$\frac{2_{\beta}\rightarrow 0_g}{2_g\rightarrow 0_g}$&rms\\
&&&&&&&$\times 10^3$&&$\times 10^3$  \\
\hline
\endhead
\hline
\endfoot
$   ^{  98}$Ru&Exp & $1.44(25)$ & $$&$ $ & $$&$1.62(61)$ & $36.0(152)$&$$&\\
 &TH&$1.50$ & $1.88$&$2.26 $ & $2.66$&$1.50$ & $0.0$&$0.25$&$0.29$&0,0477&\\
 &MR& $1.58$ & $2.10$&$2.70 $ & $3.40$&$1.58$ & $0.0$&$0.24$&$11.50$&0.0498&\\
 &M& 1.71& 2.52 &3.61& 5.27 &1.71 &0.0 &0.82& 4.24&0.0942\\
 &D& 1.82 &2.62 &3.42 &4.22& 1.82 &0.0& 1.36& 3.60&0.1420\\
 &K& 1.77& 2.81 &4.63 &8.42 &1.77& 0.0& 1.27 &27.84&0.1198 \\\\

$   ^{100}$Ru& Exp& $1.45(13)$ & $$&$ $ & $$&$0.64(12)$ & $41.1(52)$&$0.98(15)$&&\\
\hline
\end{longtable}
}

\newpage
\scriptsize{
\begin{longtable}{lllllllllllllllll}
\caption{ Comparison of experimental data \cite{data} (upper line) for several B(E2) ratios of axially symmetric prolate deformed nuclei to predictions (lower line) by the Bohr Hamiltonian with the Tietz-Hua potential (TH), Manning-Rosen potential (MR) \cite{chabab2016electric}, Morse potential (M) \cite{Inci11}, Davidson potential (D) \cite{bonat11} and Kratzer potential (K) \cite{bonat13}
\label{table4}}\\
\hline
nucl. && $\frac{4_g\rightarrow 2_g}{2_g\rightarrow 0_g}$ & $\frac{6_g\rightarrow 4_g}{2_g\rightarrow 0_g}$ &$\frac{8_g\rightarrow 6_g}{2_g\rightarrow 0_g}$ & $\frac{10_g\rightarrow 8_g}{2_g\rightarrow 0_g}$ & $\frac{2_{\beta}\rightarrow 0_g}{2_g\rightarrow 0_g}$&$\frac{2_{\beta}\rightarrow 2_g}{2_g\rightarrow 0_g}$&$\frac{2_{\beta}\rightarrow 4_g}{2_g\rightarrow 0_g}$&$\frac{2_{\gamma}\rightarrow0_g}{2_g\rightarrow 0_g}$&$\frac{2_{\gamma}\rightarrow 2_g}{2_1\rightarrow 0_g}$&$\frac{2_{\gamma}\rightarrow 4_g}{2_g\rightarrow 0_g}$&rms\\
&&&&&&$\times 10^3$&$\times 10^3$&$\times 10^3$&$\times 10^3$&$\times 10^3$&$\times 10^3$&\\
\hline
\endfirsthead
\caption{(continued)}\\
\hline
nucl. && $\frac{4_g\rightarrow 2_g}{2_g\rightarrow 0_g}$ & $\frac{6_g\rightarrow 4_g}{2_g\rightarrow 0_g}$ &$\frac{8_g\rightarrow 6_g}{2_g\rightarrow 0_g}$ & $\frac{10_g\rightarrow 8_g}{2_g\rightarrow 0_g}$ & $\frac{2_{\beta}\rightarrow 0_g}{2_g\rightarrow 0_g}$&$\frac{2_{\beta}\rightarrow 2_g}{2_g\rightarrow 0_g}$&$\frac{2_{\beta}\rightarrow 4_g}{2_g\rightarrow 0_g}$&$\frac{2_{\gamma}\rightarrow0_g}{2_g\rightarrow 0_g}$&$\frac{2_{\gamma}\rightarrow 2_g}{2_1\rightarrow 0_g}$&$\frac{2_{\gamma}\rightarrow 4_g}{2_g\rightarrow 0_g}$&rms\\
&&&&&&$\times 10^3$&$\times 10^3$&$\times 10^3$&$\times 10^3$&$\times 10^3$&$\times 10^3$&\\
\hline
\endhead
\hline
\endfoot
$^{150}$Nd&Exp & $1.52(4)$ & $1.84(14)$&$ 2.05(13)$ & $$&$4.4(8)$ & $61.7(98)$&$174(55)$&$26.1(22)$&$49.6(26)$&$14.8(98)$\\
 &TH&$1.49$ & $1.77$&$2.05 $ & $2.42$&$7.06$ & $28.56$&$199.77$&$81.16$&$125.16$&$7.27$&0,0144\\
 &MR& $1.53$ & $1.91$&$2.36 $ & $2.97$&$25.36$ & $69.51$&$349.03$&$6.25$&$136.16$&$8.07$&0.0414\\
  &K&1.55& 1.98& 2.55& 3.40& 36.4& 93.5& 443& 95.9& 150.0& 9.0&0.0664\\\\

$^{152}$Sm &Exp&$1.45(5)$&$1.70(7)$&$1.98(14)$&$2.22(25)$&$6.4(7)$&$38.2(43)$&$132(15)$&$25.1(17)$&$64.6(48)$&$5.4(5)$\\
 & TH&$1.51$ & $1.84$&$2.22$&$2.70$&$25.17$ & $62.89$&$284.15$&$69.86$&$107.33$&$6.22$&0,0085\\
 & MR&$1.51$ & $1.84$&$2.22$&$2.70$&$25.17$ & $62.89$&$284.15$&$69.86$&$107.33$&$6.22$&0.0566\\
\hline
\end{longtable}
}


 %

\bibliographystyle{IEEEtran}
\bibliography{Manuscript-TH}

\end{document}